\documentclass[10pt,twocolumn]{article}
\usepackage{graphics}

\begin{document}   

{\bf Comment on "First Order Transition in the Ginzburg-Landau Model"}

\vspace{4mm}

In a recent Letter, Curty and Beck \cite{cu-be00}
have shown very interesting results 
which indicate that the Ginzburg-Landau (G-L) transition becomes 
first order when the coherence length $\xi$=$\xi_0|t|^{-\frac{1}{2}}$ 
($t\equiv T/T_0$-1 is the reduced temperature)
becomes of the order of the lattice spacing $\varepsilon$. 
They considered the lattice G-L model parametrized by: 
$\sigma\equiv \frac{\varepsilon^2}{\xi_0^2}$, which controlls 
the strength of amplitude
fluctuations (that grow as $\sigma$ decreases), 
and $V_0\equiv \frac{1}{k_B}\frac{a}{b} \gamma$, 
which governs the overall strength of 
the complex G-L field $\psi=|\psi|\exp[i\theta]$. They treated
the model by a variational approximation 
and got a criterion for first order transition in the form of an
inequality which involves the spatial dimension $d$.
They concluded that their criterion for $d=2$ is not clear and 
that doubt remained if the first order found for this case is 
not an artifact of the used approximation.

Here I will present clear evidences that for $d$=2 a first order
transition takes place when 
$\xi$ becomes $\sim \varepsilon$ ($\sigma \sim 1$)
and that this is
connected with a sudden proliferation of vortices. 
Similar results where reported in \cite{AF99} although using a
different parametrization of the G-L model which obscures
the comparison with \cite{cu-be00}.
The $d=2$ G-L hamiltonian  
$H[\sigma,V_0]$ was 
simulated on $L \times L$ lattices. 
The measured phase diagram on the plane 
$\tilde{T}$-$\tilde{\sigma}^{-\frac{1}{2}}$ is
showed in Fig.1 where, as in ref. \cite{cu-be00}, 
$\tilde{T}=T/(-tV_0)$ and $\tilde{\sigma}=-t\sigma$. 
\begin{center}
\begin{figure}[h]
\centering
\scalebox{0.5}[0.35]{
\includegraphics{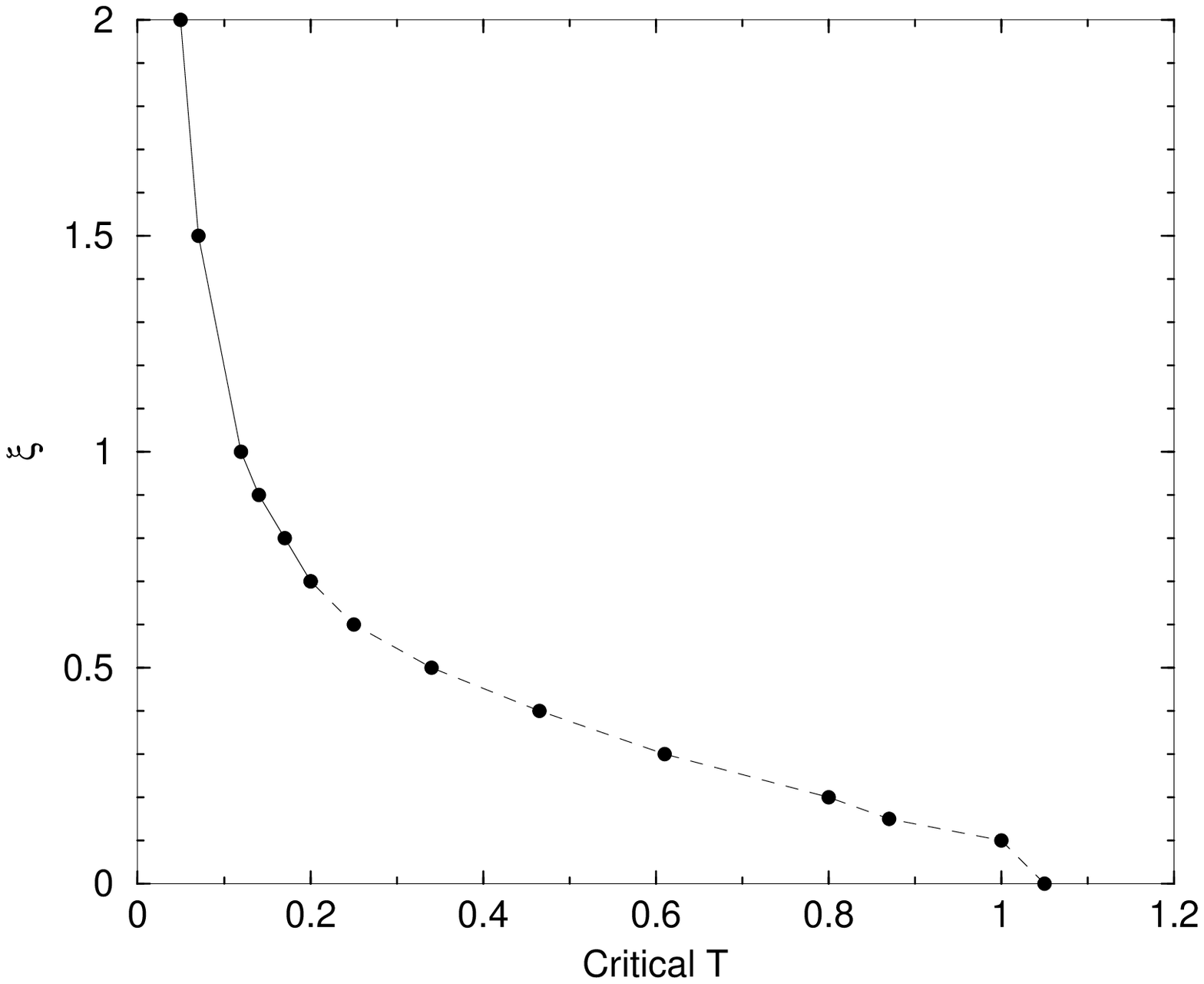}
} 
\vspace{-3mm}
\caption{Phase diagram in the $(\tilde{T},
\tilde{\sigma}^{-\frac{1}{2}})$ plane
for $L=40$.}
\end{figure}
\end{center}

\vspace{-7mm}
Above $\xi/a\equiv$ $\tilde{\sigma}^{-\frac{1}{2}}$ = 0.8 the phase 
transition line changes from second order (dashed line) 
to first order (filled line).
The double peak of the energy density $e$ histogram corresponding 
to the two coexisting phases, 
characteristic of a first-order transition, is showed in 
Fig. 2-a for
$\tilde{\sigma}^{-\frac{1}{2}}=0.85$. Both peaks remain
fixed as $L$ increases and the width of each
of them clearly scales as $\sqrt(\frac{1}{L^D})=\frac{1}{L}$, 
due to ordinary non-critical fluctuations 
(Fig. 2-b). In addition, a strong hysteresis effect was found
for $e$ when considering heating and cooling runs. 
\begin{figure}[h]
\centering
\vspace{-3mm}
\scalebox{0.45}[0.3]{
\includegraphics{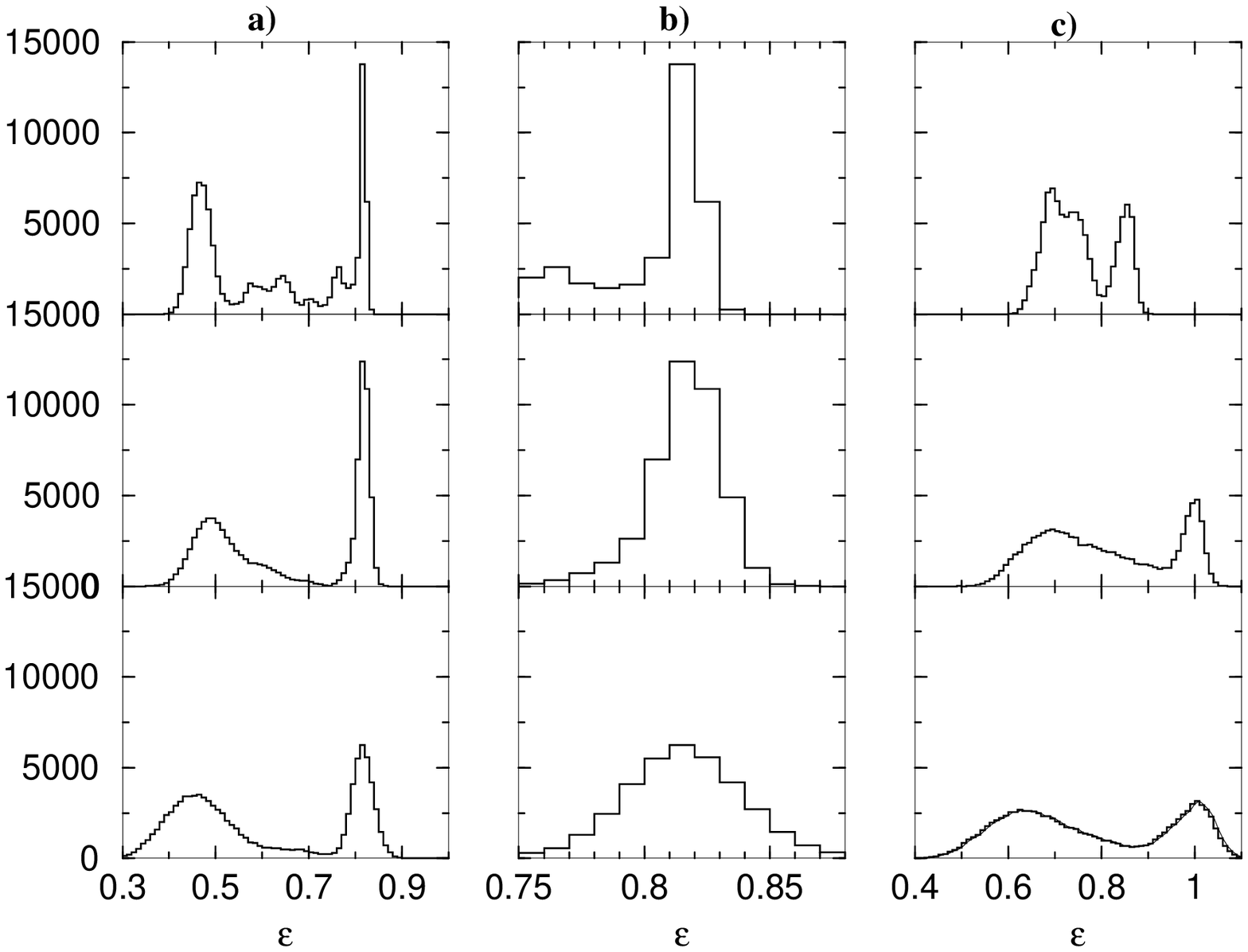}
} 
\vspace{-3mm}
\caption{Histograms of $e$ for $L$=10 (below), $L$=20 (middle) 
and $L$=40 (above). a) $\tilde{\sigma}^{-\frac{1}{2}}=0.85$. 
 b) Zoom of the right peak. 
 c) $\tilde{\sigma}^{-\frac{1}{2}}=0.75$. }
\vspace{-8mm}
\label{fig:histo}
\end{figure}
On the other hand, for $\tilde{\sigma}^{-\frac{1}{2}} \le 0.75$ 
the peaks are much lower
and wider, they move towards to an 
intermediate value of $e$ as $L$ increases and their width
do not scale as $\frac{1}{L}$ 
(Fig. 2-c). Moreover, no hysteresis in $e$ is found.

The central role played by vortex excitations
in determining the nature of the phase transition
can be seen in 
Fig. 3-a where $v$ is plotted vs. $\tilde{T}$ for different 
values of $\tilde{\sigma}$ and $L$=40.
For $\tilde{\sigma}$=1   
there is a clear discontinuity in the vortex density $v$ (Fig. 3-b). 
As long as 
$\tilde{\sigma}$ increases the jump becomes more smooth and moves 
to higher values of $\tilde{T}_c$ until for 
$\tilde{\sigma}=100$ one gets something very close
to the Kosterlitz Thouless (K-T) behavior of the XY model.
The enhancement of vortex production when amplitude fluctuations are
large is due basically to the fact that 
they decrease the energy of vortices.
\begin{figure}[h]
\centering
\scalebox{0.42}[0.23]{
\includegraphics{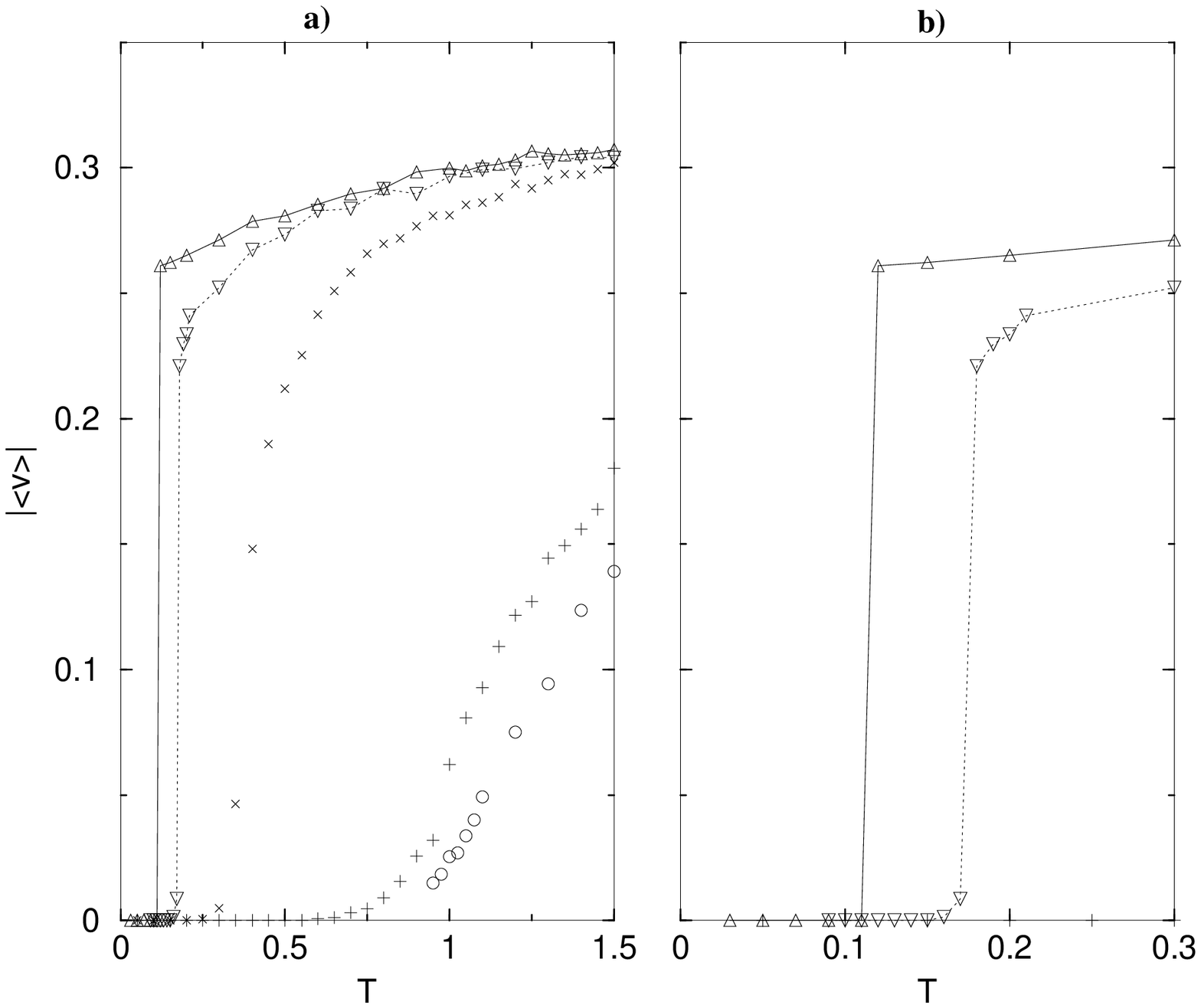}
}
\vspace{-5mm}
\caption{a) $v$ vs. $\tilde{T}$ for $\tilde{\sigma}$=1 ($\triangle$), 
$\tilde{\sigma}$=1.5625 ($\nabla$),
$\tilde{\sigma}$=4 ($\times$), $\tilde{\sigma}$=100 (+), 
XY model(o). b) Zoom of \ref{fig:allvortices}-(a)}
\vspace{-2mm}
\label{fig:allvortices}
\end{figure}

Therefore, in the G-L model the nature of the phase 
transition depends dramatically on the value of 
$\tilde{\sigma}$: For $\tilde{\sigma}<$1.5625 
($\frac{\xi}{\varepsilon} > 0.8$)
the density of vortices experiments a discontinuous jump which 
coincides with a first order transition. 
On the other hand, for $\tilde{\sigma}\gg 1$ the G-L reduces
to the XY model with the more subtle 
K-T phase transition.

\vspace{2mm}

Hugo Fort

Instituto de F\'{\i}sica, Facultad de Ciencias

Igu\'a 4225, 11400 Montevideo, Uruguay


\end{document}